\documentclass[a4paper]{jpconf}
\usepackage{amsmath}
\usepackage{graphicx}	       % EPS-Grafiken einbinden
\usepackage[english]{babel}     % Babel-Paket laden
\begin{document}

\title{Spherical Orbifolds for Cosmic Topology}
\author{Peter Kramer}
\address{Institut f\"ur Theoretische Physik der Universit\"at \\
T\"ubingen, Germany}
\ead{peter.kramer@uni-tuebingen.de}

%arxiv:1201.1875\\
%PACS: 95.85.Bh, 98.80.-k, 02.40.Pc, 61.50.Ah

\begin{abstract}
Harmonic analysis is a tool to infer cosmic topology from the measured astrophysical cosmic microwave background CMB radiation.
For overall positive curvature, Platonic spherical manifolds are  candidates for 
this analysis. We combine  the specific point symmetry of the Platonic manifolds 
with their deck transformations. This analysis in topology leads from manifolds to  orbifolds. We discuss 
the deck transformations of the orbifolds and give eigenmodes for the harmonic analysis as linear combinations of Wigner polynomials on the 3-sphere. These provide  new tools for detecting  cosmic topology from the CMB radiation.
\end{abstract}

\section{Motivation and elements of topology.}
\label{Mo}

\begin{figure}[b]
\begin{center}
\includegraphics[width=0.57\textwidth]{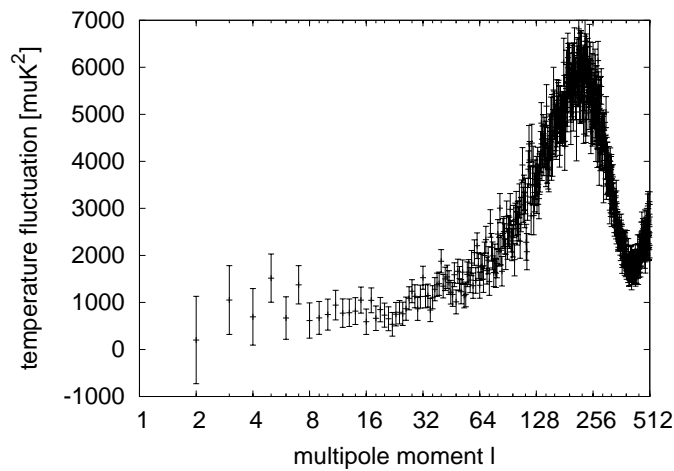}
\end{center}
\caption{\label{fig:Fig1}
CMB Amplitude from the WMAP team \cite{LAR11} as a function of the multipole order $l$.}
\end{figure}

Data observed by the Wilkinson satellite \cite{LAR11} and  expected from the Planck satellite display the black-body cosmic microwave background  CMB radiation with temperature T= 2.75 K and 
$\lambda_{\rm max}$=19 mm, Fig. \ref{fig:Fig1}. This radiation is generally assumed to stem from the big bang.
A small amplitude 
for low CMB multipoles $l$ suggest selection rules. Do they arise from cosmic topology?

\begin{figure}[t]
\begin{center}
\includegraphics[width=0.4\textwidth]{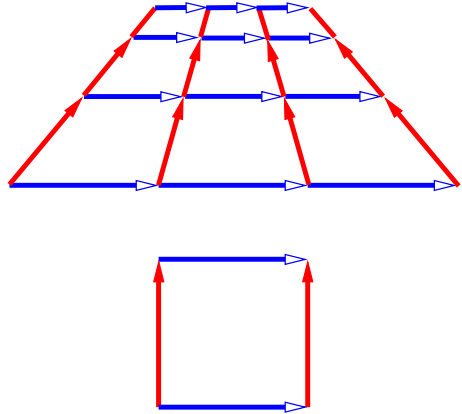}
\end{center}
\caption{\label{fig:Fig2}
{\bf Parallel universes?}  Lift the square to its universal cover. By matching with horizontal/vertical neighbours we  cover the plane by parallel squares.} 
\end{figure}

\begin{figure}[b]
\begin{center}
\includegraphics[width=0.8\textwidth]{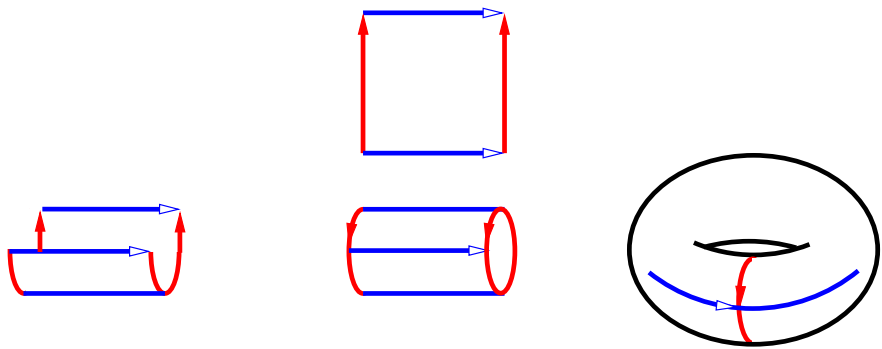}
\end{center}
\caption{\label{fig:Fig3}
{\bf Folded universe?} Glue the square by local homotopy.  
Horizontal edge folding  plus gluing yields the cylinder.
Successive vertical edge folding plus gluing gives the 2-torus $T^2$.}
\end{figure}

In this first section  we introduce  topology and functional analysis for  manifolds. Multiply connected spherical 3-manifolds can model  
the space part of cosmic space-time. 
We demonstrate topology in two dimensions from a square. First in Fig. \ref{fig:Fig2} we lift this square to the Euclidean plane $E^2$, called its universal cover.
The square becomes part of a square tiling of $E^2$.
The group that generates the tiling is called the deck group.
In a second approach, Fig. \ref{fig:Fig3},  we fold and glue the square into the torus. The gluings correspond to  a group, termed the homotopy group $H$ of the square. 
Homotopy and covering are linked by a theorem of Seifert and Threlfall \cite{SE34} which says that the homotopy and the deck group are isomorphic.  Fig. \ref{fig:Fig3} illustrates the gluings, but may be misleading with respect to curvature: The square is Euclidean and shares
zero curvature with its cover, the Euclidean plane. So the two views of a topological manifold,  in terms of its universal cover or
its self-gluing, allow to  express different topological aspects and must be taken in conjunction.

\begin{figure}[t]
\begin{center}
\includegraphics[width=0.4\textwidth]{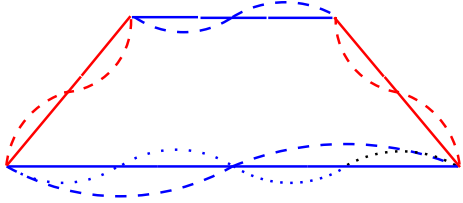}
\end{center}
\caption{\label{fig:Fig4} Periodic functions on the square from Fourier analysis.}
\end{figure}

\subsection{Functions carried on a topological manifold.}
\label{Fu}
We pass to functions carried by the  square. On the covering plane, periodic boundary conditions on the square tiles  yield twofold Fourier analysis. Any Fourier basis function  
\begin{eqnarray}
\label{or1}
\psi_{n_1,n_2} (x,y)=\exp i(k_1x+k_2y),\; k_1L=2\pi n_1,\; k_2L=2\pi n_2,
\; \; (n_1, n_2) = 0, \pm 1, \pm2,...
\end{eqnarray}
is invariant under the deck group $H=C_{\infty} \times C_{\infty}$. As any  deck group, $H$  must act without fixpoints,
compare Fig. \ref{fig:Fig2}.

A random function on a topological manifold
is generated by a random superposition of its characteristic eigenmodes. In cosmic topology one compares  such random 
superpositions with the observed CMB radiation. A similarity of both is thought to indicate the physical topology of the 
space part from space-time. 

Before a comparison can be  made,  we must pass from the 2-dimensional square to 3-dimensional and more realistic manifolds.
In contrast to zero curvature found for the square,  positive curvature suggests to study spherical manifolds to which we turn in section \ref{Spher}.

The exploration of point symmetry for topology will lead us to the notion of orbifolds. 
In contrast to topological manifolds, orbifolds have deck group actions with singular fixpoints, that is,   fixpoints of non-zero but finite order.

\subsection{Einstein's field  equations.}
\label{Ei}
Einstein's field equations determine the metric structure of space-time from the energy-momentum tensor.
\begin{eqnarray}
\label{or2}
&& (R_{ik}-\frac{1}{2} g_{ik})-\lambda\: g_{ik}=-\kappa\: T_{ik},
\\ \nonumber
&&\kappa = \frac{8\pi C}{c^4}= 23.07\times  10^{-48} \frac{s^2}{g\:cm}.
\end{eqnarray}
Here, $R_{ik}$ is the Ricci tensor,
$g_{ik}$ the metrical tensor,
$T_{ik}$ the energy-momentum tensor,
$C$ Newton's  gravitational, 
$\lambda$ the cosmological constant. 
The Ricci tensor contains second derivatives of  the metrical tensor, and so the field equations are second order partial differential equations for   the metrical tensor $g_{ik}$ in terms of a given  energy-momentum tensor $T_{ik}$.
Note: The cosmic mass distribution is smoothed out into a mass fluid!

Einstein in his first paper \cite{EI17}, as is also explained in \cite{LA56}, applied his  equations to space taken as a 3-sphere.
This choice assumes positive average curvature and closure, and still  is relevant  for   present-day  cosmic topology.

\subsection{Selection rules from point symmetry and orbifolds.}
\label{Se}
Angular coordinates with respect to  the midpoint of the square are
\begin{equation}
\label{or3}
 z= \rho\: \exp(i\phi),\: \rho \geq 0,\: 0\leq \phi <2\pi.
\end{equation}
Functions in multipole form may be written as 
\begin{equation}
\label{or4}
 f(z) = R_m(\rho) \exp(im\phi), m=0,\pm 1,\pm2,...
\end{equation}
Selection rules from point symmetry  under ther cyclic group $C_4$ then imply
\begin{equation}
\label{or5}
m \equiv\: 0\:   {\rm mod}\:  4=\: 0,\pm 4, \pm 8,...
\end{equation}

\begin{figure}[t]
\begin{center}
\includegraphics[width=0.6\textwidth]{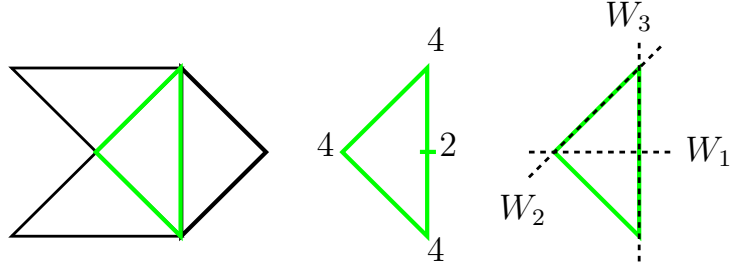}
\end{center}
\caption{\label{fig:Fig5}
{\bf Orbifolds}: The square is an   orbit under the cyclic group $C_4$ of the triangular orbifold $\Delta$.
Left: Three $\Delta's$ cover edges of $\Delta$.
Middle: Singular fixpoints of finite orders $(4,4,4,2)$ on the edges $\partial \Delta$ of the orbifold $\Delta$. 
Right: Three Weyl reflections $W_1, W_2, W_3$ in the dashed lines generate a Coxeter group $\Gamma$. Its rotational unimodular subgroup
$S\Gamma: \langle W_1W_2,W_2W_3, W_3W_1\rangle $ of even reflections generates a triangular tiling of the plane into orbifolds, preserves orientation, and yields selection rules. The unimodular subgroup $H=S\Gamma$ is the deck and homotopy group of the  orbifold $\Delta$!}
\end{figure}

The introduction of point symmetry into topology leads to orbifolds rather than manifolds.
For the square we explain the connection in Fig. \ref{fig:Fig5}. A triangle $\Delta$ forms  a fundamental domain 
for the cyclic group $C_4$ acting on the square. The application of even products from a Coxeter group to a single preimage
covers the plane $E^2$ by copies of the triangle. This action has rotational fixpoints on edges of the triangle, marked in Fig. \ref{fig:Fig5}, and so is not fixpoint-free as required for the deck transformation of a topological manifold. Instead, the triangle with its fixpoints forms an orbifold.

\section{Spherical spaces.}
\label{Spher}
After the introductory section \ref{Mo} we turn now to realistic models of cosmic 3-space.
For these  models of cosmic space we must extend all the notions illustrated on the square to appropriate manifolds. 
The spaces should  have  dimension 3 and should have  positive curvature like  spherical spaces.
Such spaces result by closing pieces from the 3-sphere as used by Einstein \cite{EI17} for his initial cosmology.
Our exposition now necessarily becomes more technical and uses results given in the references.
Harmonic analysis  on the manifolds  aims at  explaining 
the observed low amplitudes at small multipole order $l$  of the cosmic microwave background. We analyze assumptions of point symmetry  and randomness  for Platonic spherical spaces. There emerge four new spaces, called orbifolds, with low volume fraction from the 3-sphere  and sharp multipole selection rules in their eigenmodes.

\begin{figure}[t]
\begin{center}
\includegraphics[width=0.62\textwidth]{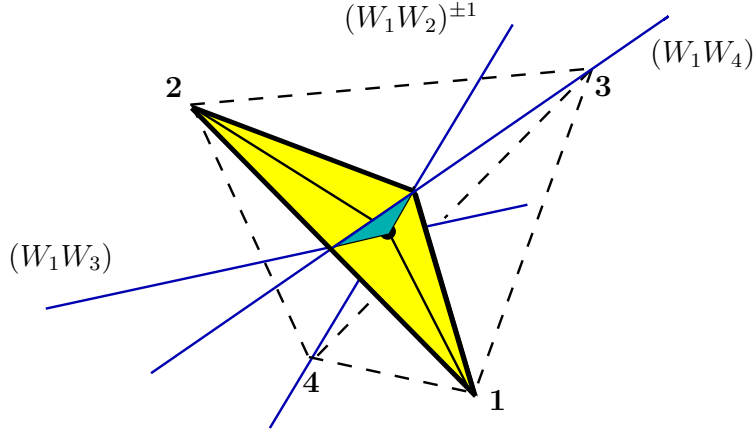} 
\end{center}
\caption{\label{fig:Fig6}
{\bf Tetrahedral orbifold.} Face gluing deck rotations of the tetrahedral  duplex orbifold (yellow) $N8$ in Euclidean version. 
The vertices of the Platonic tetrahedron are marked by the  numbers $(1, 2, 3, 4)$. A small tetrahedron of volume fraction 
$\frac{1}{12}$ inside the Platonic tetrahedron 
is the orbifold (yellow colour).
The four covering  rotations  $(W_1W_3), 
(W_1W_2)^{\pm 1},(W_1W_4)$ of this orbifold, corresponding to the covering of its four faces,  are products of Weyl reflections from the Coxeter group with diagram $\Gamma =\circ - \circ -\circ -\circ$. They  are marked 
by their rotation axes, intersection lines of pairs of Weyl reflection planes. The three intersections of these axes with the orbifold
form the edges of the glue triangle (light blue area) of two Coxeter simplices. The  inner points of its edges  have  the orders (2, 3, 2) 
of their covering rotations.}
\end{figure}

\section{Eigenmodes for spherical spaces.}
\label{Eig}
Spherical spaces or spherical space forms are abstractly described as the quotients of the covering  manifold $S^3$ by the deck group $H$ of order $|H|$ acting fixpoint-free.
The topologies  are then classified as space forms ${\cal M}=S^3/H$. They were characterized with faithful representations of groups $H$ by Wolf in \cite{WO84} pp. 198-230. This characterization leaves open  the geometry of the space, and so gives no access to point symmetry. Therefore we prefer a  fully  geometric approach \cite{KR10}, starting from homotopies  derived by Everitt  \cite{EV04}.

The 3-sphere $S^3$ of constant positive curvature $\kappa=1$, Einstein 1917 \cite{EI17}, covers all spherical  space forms.
There is a one-to-one correspondence  of $S^3$ and $SU(2,C)$ given by
\begin{eqnarray}
\label{or6}
&&  x=(x_0,x_1,x_2,x_3),\\ \nonumber
&&\leftrightarrow   u(x)=\left[ \begin{array}{ll}
           x_0-ix_3&-x_1-ix_2\\
           x_1-ix_2& x_0+ix_3\\
          \end{array} \right]\: \in SU(2,C),\\ \nonumber 
&&x_0^2+x_1^2+x_2^2+x_3^2={\rm det}(u)=1.
\end{eqnarray}
The rotation group $SO(4,R)$ acts on the points of the 3-sphere by the correspondence 
\begin{equation}
\label{or7}
 SO(4,R)= (SU^l(2,C) \times SU^r(2,C))/\pm(e,e),
\end{equation}
and its elements $(g_l,g_r)$ transform these points  by 
\begin{equation}
\label{or8}
(g_l,g_r): u \rightarrow g_l^{-1}ug_r.
\end{equation}
Wigner introduced the matrices $D^j_{m_1m_2}(u)$ as the irreducible representation matrices of the group $SU(2,C)$ \cite{ED57}. 
Considered as  homogeneous polynomials in the four elements of the matrix $u(x)$, we call them Wigner polynomials. Their overall degree is
$2j$ and is conserved under the linear transformations eq. \ref{or9}.
The Wigner polynomials  are a complete and orthonormal set of functions on the 3-sphere $S^3$.
Moreover they are harmonic \cite{KR10}, that is,  vanish under the Laplacian on the Euclidean space $E^4$ 
that embeds the 3-sphere.

\begin{figure}[t]
\begin{center}
\includegraphics[width=0.7\textwidth]{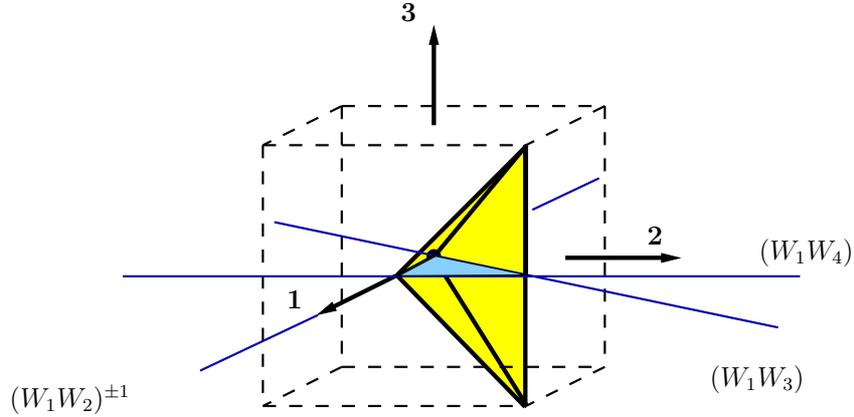} 
\end{center}
\caption{\label{fig:Fig7}
{\bf Cubic orbifold}. Face gluing deck rotations of the cubic duplex orbifold (yellow) $N9$ in Euclidean version. Its four covering rotations  $(W_1W_3),  
(W_1W_2)^{\pm 1}, (W_1W_4)$, products of pairs of  Weyl reflections from the Coxeter group with diagram $\Gamma =\circ \stackrel{4}{-} \circ -\circ -\circ$, are  marked 
by rotation axes, intersection lines of pairs of  Weyl planes. The three intersections of these axes with the orbifold
form the edges of the glue  triangle (light blue area). Its inner edge points have the orders (2, 4, 2) 
of their covering rotations.}
\end{figure}

The action of $SO(4,R)$ on Wigner polynomials has the form
\begin{eqnarray}
\label{or9}
 &&(T_{(g_l,g_r)}D^j_{m_1,m_2})(u)=D^j_{m_1m_2}(g_l^{-1}ug_r)
\\ \nonumber 
&&= \sum_{m_1'm_2'} D^j_{m_1'm_2'}(u) D^j_{m_1m_1'}(g_l^{-1}) D^j_{m_2'm_2}(g_r), 
\end{eqnarray}
and so is determined by pairs of  representation matrices of the group $SU(2,C)$.
 By eq. \ref{or9} we can take advantage of the elaborate techniques of angular momentum
from \cite{ED57}.

Each spherical manifold carries a specific harmonic  basis of eigenmodes, invariant under a deck group $H$, and obeying homotopic boundary conditions.

The eigenmodes of a spherical manifold offer two alternative views  on different domains: 
(i) On a single closed manifold ${\cal M}$, they  allow to expand square integrable observables. These obey homotopic boundary conditions for their functional values on faces and edges.\\ 
(ii) The covering 3-sphere is 
tiled by copies of ${\cal M}$. Any
eigenmode now must on each different tile repeat its value, and fulfill the homotopic 
boundary conditions. It follows that the eigenmodes must display selection rules in comparison to a general polynomial basis on the 3-sphere. This second view allows for  a comparison of observables for different topologies on the same domain, Einstein's 3-sphere.

In topology, these two views present  (i) the local concept of homotopy of ${\cal M}$, and 
(ii) the concept of deck transformations from  the group $H={\rm deck}({\cal M})$ that generates the tiling on the  3-sphere as universal cover. 
Seifert and Threlfall \cite{SE34} pp. 195-8 prove the equivalence of the two views: the fundamental or first homotopy group  $\pi_1({\cal M})$ that generates all the gluings is isomorphic to the group of deck transformations ${\rm deck}({\cal M})$ that generates the tiling on the cover.

\section{Spherical Platonic 3-manifolds.}
\label{Spe}

\begin{figure}[t]
\begin{center}
\includegraphics[width=0.6\textwidth]{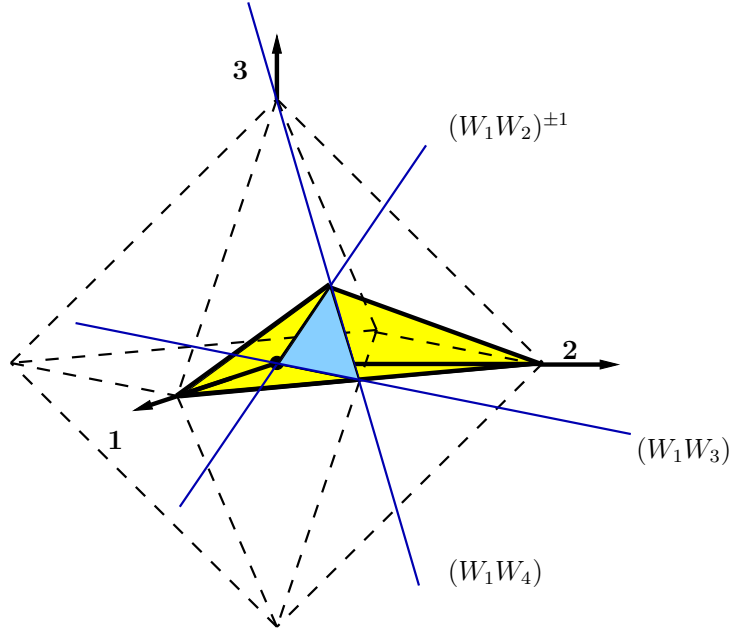} 
\end{center}
\caption{\label{fig:Fig8}
{\bf Octahedral orbifold}. Face gluing deck rotations of the octahedral duplex
orbifold (yellow) $N10$ in Euclidean version. The corresponding  four products $ (W_1W_3), 
(W_1W_2)^{\pm 1}, (W_1W_4)$ of pairs of Weyl reflections from $\Gamma =\circ - \circ  \stackrel{4}{-}\circ -\circ$ are marked 
by rotation axes, intersection lines of pairs of  Weyl reflection planes. The three intersections of these axes with the orbifold
form the  edges of the glue  triangle (light blue area). Its inner edge points have the orders (2, 3, 2) 
of their covering rotations.}
\end{figure}

We study  here the family of Platonic spherical 3-manifolds.
For each Platonic polyhedron we construct in \cite{KR10} on the 
universal cover, the 3-sphere, a unique group $H$  of fixpoint-free deck transformations 
acting on the 3-sphere 
as a subgroup of a Coxeter group $\Gamma$ . 
By the theorem from \cite{SE34}, the deck groups $H$ are isomorphic to, and were derived in \cite{KR10} from,  the fundamental or first homotopy groups constructed by Everitt \cite{EV04}.

\begin{table}[t]
$
\begin{tabular}{|l|l|l|l|l|l|}\hline
Coxeter diagram $\Gamma$ & $|\Gamma|$ & polyhedron ${\cal M}$ & $H={\rm deck}({\cal M})$ & $|H|$ & Reference \\ \hline
$\circ -\circ -\circ - \circ$               & $120$  & tetrahedron $N1$ & $C_5$         & $5$ & \cite{KR08} \\  \hline
$\circ \stackrel{4}{-} \circ -\circ -\circ$ & $384$  & cube $N2$        & $C_8$         & $8$ & \cite{KR09} \\                                                           &        & cube $N3$        & $Q$           & $8$ &\cite{KR10} \\  \hline
$\circ -\circ \stackrel{4}{-}\circ - \circ$ & $1152$ & octahedron $N4$  & $C_3\times Q$ & $24$ &\cite{KR10} \\
                                            &        & octahedron $N5$  & $B$           & $24$ &\cite{KR10} \\
                                            &        & octahedron $N6$  & ${\cal T}^*$  & $24$&\cite{KR10} \\ \hline
$\circ \stackrel{5}{-} \circ -\circ  -\circ$  & $120^2$  & dodecahedron$N1'$& $ {\cal J}^*$ & $120$&\cite{KR05}\\
\hline
\end{tabular}
$
\caption{\label{table:Table1}
4 Coxeter groups $\Gamma$, 4 Platonic polyhedra ${\cal M}$, 7 deck groups $H={\rm deck}({\cal M})$ of order $|H|$.
$C_n$ denotes a cyclic, $Q$ the quaternion, ${\cal T}^*$ the binary tetrahedral, ${\cal J}^*$ the binary icosahedral group. The symbols $Ni$ are taken or extended  from  \cite{EV04}. The last column gives the reference for the harmonic analysis. The diagram for the icosahedral Coxeter group is 
corrected compared to  \cite{KR10}.}
\end{table}

\begin{table}[t]
\begin{equation*}
\begin{array}{|l|l|l|l|l|} \hline
\Gamma & a_1 & a_2 &a_3& a_4\\ \hline
\circ -\circ -\circ - \circ &  (0,0,0,1)&(0,0,\sqrt{\frac{3}{4}},\frac{1}{2})
& (0,\sqrt{\frac{2}{3}},\sqrt{\frac{1}{3}},0)& (\sqrt{\frac{5}{8}},\sqrt{\frac{3}{8}},0,0)\\ \hline
\circ \stackrel{4}{-} \circ -\circ -\circ&  (0,0,0,1)& (0,0,-\sqrt{\frac{1}{2}},\sqrt{\frac{1}{2}})
&(0,\sqrt{\frac{1}{2}},-\sqrt{\frac{1}{2}},0)&(-\sqrt{\frac{1}{2}},\sqrt{\frac{1}{2}},0,0)\\ \hline
\circ -\circ \stackrel{4}{-}\circ - \circ& (0,\sqrt{\frac{1}{2}},-\sqrt{\frac{1}{2}},0)&(0,0,-\sqrt{\frac{1}{2}},\sqrt{\frac{1}{2}})
&(0,0,0,1)& (\frac{1}{2},\frac{1}{2},\frac{1}{2},\frac{1}{2})\\ \hline
\circ  \stackrel{5}{-} \circ -\circ  -\circ& (0,0,1,0)&(0,-\frac{\sqrt{-\tau+3}}{2},\frac{\tau}{2},0)
&(0,-\sqrt{\frac{\tau+2}{5}},0,-\sqrt{\frac{-\tau+3}{5}})&(\frac{\sqrt{2-\tau}}{2},0,0,-\frac{\sqrt{\tau+2}}{2}) \\ \hline
\end{array}
\end{equation*}
\caption{\label{table:Table2}
The Weyl vectors $a_s$  
for the four  Coxeter groups $\Gamma$ from Table \ref{table:Table1}  with  $\tau:=\frac{1+\sqrt{5}}{2}$.}
\end{table}

\begin{table}[t]
$
\begin{tabular}{|l|l|l|l|l|l|}\hline
Coxeter         & $|H|$   & polyhedron &point group     & deck generators &fixpoint \\ 
diagram $\Gamma$&          &and orbifold     &$M$,$|M|$ &&order k
\\ \hline 
$\circ -\circ -\circ - \circ$               & $5 $  & tetrah N8 & $A(4), 12$ & $(W_1W_3), (W_1W_2)^{\pm 1}, (W_1W_4)$
&(2, 3, 2)
\\  \hline
$\circ \stackrel{4}{-} \circ -\circ -\circ$ & $8$  & cube N9 & $O, 24$ & $(W_1W_3), (W_1W_2)^{\pm 1}, (W_1W_4)$ 
&(2, 4, 2) 
\\  \hline
$\circ -\circ \stackrel{4}{-}\circ - \circ$ & $24$ & octah N10 &$O$, 24 & $(W_1W_3), (W_1W_2)^{\pm 1},  (W_1W_4)$
&(2, 3, 2)\\ \hline
$\circ\stackrel{5}{-}  \circ - \circ - \circ $ & $120$ & dodec N11 &${\cal J}$, 60 
& $(W_1W_3), (W_1W_2)^{\pm 1},  (W_1W_4)$&(2, 5, 2)
\\ \hline
\end{tabular}
$
\caption{\label{table:Table3}
From four Coxeter groups $\Gamma$, $|S\Gamma|=|M| \cdot |H|$ and Platonic polyhedra to orbifolds. Point groups $M$, 
deck generators of orbifolds, and selected  orders $k$ for fixpoints on edges of the glue triangle of four  duplex orbifolds $N8, N9, N10, N11$, see Figs. \ref{fig:Fig6} -\ref{fig:Fig9}. $A(4) $ is the tetrahedral, $O$ the cubic, ${\cal J}$ the icosahedral rotation group. The center of the polyhedron has order $|M|$, inner points of the orbifolds are regular. Note that the Weyl reflections $W_l$ depend on the Coxeter group chosen, see Tables \ref{table:Table1}, 
\ref{table:Table2}.}
 
\end{table}

Coxeter groups $\Gamma$ \cite{CO73}, \cite{HU90} will become a main tool of the following  analysis.
They  are generated   on Euclidean 4-space with coordinates $x=(x_0, x_1, x_2, x_3)$, and on the 3-sphere, by four involutive Weyl reflections $(W_1, W_2, W_3, W_4)$ in hyperplanes, see Table \ref{table:Table2}. A reflection hyperplane $W_i$ containing the origin is characterized  by a unit vector $a_i$ perpendicular to it.
The Coxeter diagram encodes the group relations between the Weyl reflections, associated to  its four nodes.
A line connecting two nodes implies $(W_iW_{i+1})^3=e$, 
a connecting line with integer superscript  $k > 3$ implies $(W_iW_{i+1})^k=e$,
Weyl reflections for nodes not connected by lines commute with one another.
Tables \ref{table:Table1}, \ref{table:Table2}  review the relation of the Platonic polyhedra to Coxeter groups.
In Table \ref{table:Table2}  we list the four unit vectors $a_j \in E^4$, perpendicular to the Weyl 
reflection hyperplanes, for each Coxeter group.
The Coxeter group tiles the 3-sphere into $|\Gamma|$ Coxeter simplices. The initial Platonic polyhedron consists of
those Coxeter simplices which share a vertex at the center $x=(1,0,0,0)$.  
In topology we prefer  orientable manifolds. In the Coxeter groups this means that
we must restrict our attention to the subgroups generated by an even number of Weyl reflections. We call the
corresponding subgroups $S\Gamma$, where $S$ stands for unimodularity of the defining representation 
on the Euclidean space $E^4$. The order of these subgroups is $|S\Gamma|=|\Gamma|/2$. 
A set of  generators of all unimodular Coxeter groups with four Weyl reflections is given by 
\begin{equation}
\label{or10}
 S\Gamma: \{ (W_1W_2), (W_2W_3), (W_3W_4) \}.
\end{equation}
The first two products of reflection generators generate the point group $M$ of the Platonic polyhedron, the last product 
enters the group of deck operations of the Platonic tiling of the 3-sphere. 
Note that any product $(W_iW_j)$ leaves invariant the intersection of the two Weyl reflection hyperplanes 
perpendicular to the vectors $\{a_i, a_j\}$. On the 3-sphere this intersection is the geodesic rotation axis for the 
rotation $(W_i W_j)$. The representation of any such product is derived  from their Weyl reflection vectors in  \cite{KR08} eq. (60):  For two  Weyl reflection operators with Weyl unit vectors $\{a_i, a_j\}$, define with eq. \ref{or6}
\begin{equation}
 \label{or25}
v_i:=u(a_i),\; v_j:=u(a_j).
\end{equation}
Then the rotation operator for the product $(W_iW_j)$ is given by inserting $(g_l, g_r)=(v_iv_j^{-1}, v_i^{-1} v_j)$  into 
eq. \ref{or9},

\begin{equation}
 \label{or26}
T_{(W_iW_j)}= T_{(v_iv_j^{-1}, v_i^{-1} v_j)}. 
\end{equation}

As an example of an eigenmode we give from \cite{KR10} the selection rules for  the Platonic spherical cube manifold $N3$. The 3-sphere 
$S^3$ is tiled into eight spherical cubes. This tiling is the 8-cell. The deck transformations  
$H= Q$ from \cite{KR10} form the quaternion group $Q$ generated by the quaternions  $\langle \pm  e,\pm i,\pm j,\pm k\rangle$. 
The projector $P^0_{Q}$ to the identity representation of $H=Q$ applied to a Wigner polynomial is found to be 
\begin{eqnarray}
\label{or11}
&&(P^0_{Q}\:D^j_{m_1m_2})(u)=\frac{1}{8}\left[1+(-1)^{2j}\right]\left[1+(-1)^{m_1}\right]
\\ \nonumber
&&\times \left[D^j_{m_1m_2}(u)+(-1)^{m_1}D_{-m_1m_2}(u)\right],
\end{eqnarray}
it selects $j={\rm integer},m_1={\rm even}$.

\begin{figure}[t]
\begin{center}
\includegraphics[width=0.65\textwidth]{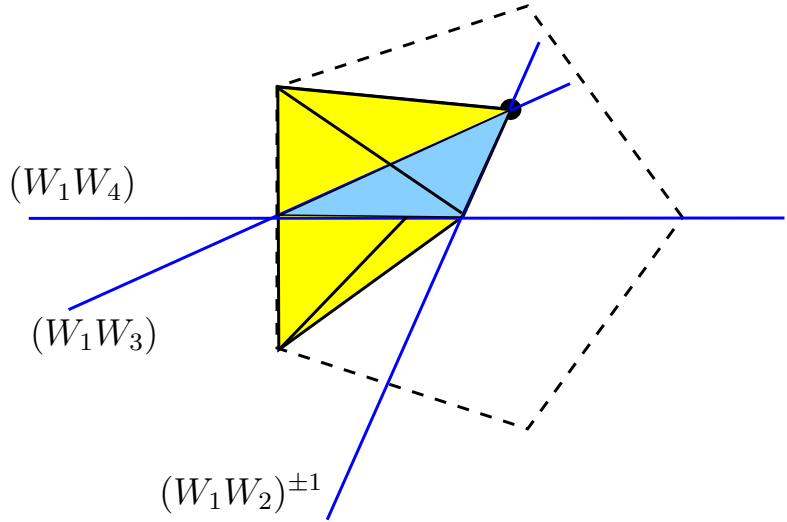} 
\end{center}
\caption{\label{fig:Fig9}
{\bf Dodecahedral orbifold}. Face gluing deck rotations of the dodecahedral duplex orbifold (yellow) $N11$, attached to a 
dodecahedral face and the center, in Euclidean version. Only one face and the center are shown. The corresponding  four covering products $ (W_1W_3), 
(W_1W_2)^{\pm 1}, (W_1W_4)$ of two Weyl reflections from 
$\Gamma =\circ  \stackrel{5}{-}\circ -\circ  -\circ $ are marked 
by rotation axes,  intersection lines of pairs of  Weyl planes. The three intersections of these axes with the orbifold
form the edges of the glue triangle (light blue area). Its  inner edge points have the orders (2, 5, 2) 
of their covering rotations.}
\end{figure}

\section{Duplices under the point group tile a polyhedron.}
\label{Dup}

In \cite{KR10} we introduce  in addition  for random functions  the notion of random symmetry under 
the point group $M$ of these manifolds. We argue there that the values of a proper random function on a polyhedral manifold 
whose shape displays the point symmetry $M$ should be independent of operations from $M$. Therefore one should explore $M$ for the analysis of the CMB radiation. The point groups are the tetrahedral group $A(4)$ for the tetrahedral 3-manifold, the cubic group $O$ 
for the cubic and octahedral 3-manifolds, and the icosahedral group ${\cal J}$ for the spherical 
dodecahedron. These point groups are unimodular subgroups w.r.t the 2-sphere 
\begin{equation}
\label{or12}
x_1^2+x_2^2+x_3^2=1,
 \end{equation}
and may be characterized on 3-space by unimodularity $S$ and by Coxeter subdiagrams as
\begin{equation}
\label{or13}
A(4)=S(\circ-\circ-\circ),\: O=S(\circ\stackrel{4}{-}\circ-\circ)\sim S(\circ-\circ\stackrel{4}{-}\circ),
\:{\cal J}=S(\circ \stackrel{5}{-} \circ-\circ).
\end{equation}

The  action of the point group $M$ on the  Platonic polyhedron with center at $x=(1,0,0,0)$ can now be decomposed into a  pre-image  and its images under $M$. The shape of a fundamental domain is not unique, but we can choose it  in compact, convex and polyhedral form: Each fundamental domain for $M$  we take as a duplex inside the proto-polyhedron, formed by gluing a Coxeter simplex and its
mirror image  under reflection in one simplex face.  This simplex face we call the glue triangle. It is shown in Figs. 
\ref{fig:Fig6}, \ref{fig:Fig7}, \ref{fig:Fig8}, \ref{fig:Fig9}.

{\bf Prop 1: Fundamental domains under point groups}:
For each Platonic polyhedron, a duplex fundamental domain of their point group $M$ may be chosen as shown 
in Figures \ref{fig:Fig6}, \ref{fig:Fig7}, \ref{fig:Fig8}, \ref{fig:Fig9}.  All of them are again spherical simplices  with four faces, have one vertex at the center of the 
polyhedron, and tile the polyhedron. 

\section{Orbifolds under the group $S\Gamma$ tile the 3-sphere.}
\label{Orb}

The Platonic 3-manifolds under their deck group $H$  in turn tile the 3-sphere into $|H|$ copies of the Platonic polyhedron as prototile. 
These tilings into Platonic polyhedra are the $m$-cells, $m=|H|$, described in \cite{SO58} and \cite{KR10}.
By composing the polyhedra, tiled into duplices, as tiles of the $|H|$-cell  tilings, it follows that the 3-sphere is tiled into $|S\Gamma|=|M| \cdot |H|$ duplices. 
With respect to actions on the 3-sphere, the Coxeter duplices, which originated as fundamental domains of the point group acting on the Platonic prototile, now become  the 
fundamental domains of the much bigger groups $S\Gamma$. In the mathematical community they are called  orbifolds. In Euclidean 
crystallography with space groups they are familiar under the name  asymmetric units. A planar Euclidean example is provided in 
Fig. \ref{fig:Fig5}. 

\subsection{The action of deck and point groups.}
The subgroup $H < S\Gamma$  of deck transformations for a given 3-manifold  acts on the 3-sphere without fixpoints.
The point group $M < S\Gamma$  by definition preserves the center of the polyhedron.  It  follows \cite{KR10B} that the intersection 
of these two subgroups contains only the identity,  $H \cap M=e$. 

Any image of the orbifold as proto-duplex and pre-image under $S\Gamma$ on the 3-sphere has a unique factorized  address, composed of a unique point group element $p\in M$ acting on the initial polyhedron, followed by a  unique deck transformation $h\in H$ from  the initial polyhedron to an image on the $|H|$-cell tiling. This leads to the following general conclusion on the group structure:

{\bf Prop 2: Unimodular Coxeter groups are products of subgroups }: Given a prototile duplex, its image tile in the Coxeter duplex tiling  of the 3-sphere results from the action of a  group element from $S\Gamma$, uniquely factorized into a deck transformation 
from $h\in S\Gamma$ and a point transformation $p \in M$.  This means that the 
elements  $g \in S\Gamma$ obey the unique subgroup product law 

\begin{equation}
\label{or14}
g\in S\Gamma:\: g= h  p,\:  h \in H,\: p \in M,\: H \cap M=e,\: |S\Gamma|=|H|\cdot  |M|.
\end{equation}
{\bf Proof}: The group $S\Gamma$ with subgroups $H,\: M$ all have faithful (one to one)  representations, \cite{WO84} p. 138, on $E^4$.
This implies  that any image under $g \in S\Gamma$ of the orbifold as proto-duplex has a unique 
factorized address as a product $g=h p,\:  h\in H, p\in M$. Similarly one can construct for the same image 
a unique factorized address $g=p' h'$.

This  product structure $S\Gamma=H \cdot M$ differs from   a direct or semidirect product. The two subgroups $H, M$ do not commute, and  the group $H$ in general is not invariant under conjugation with elements from $M$,  except in  
case of the cubic Coxeter group, analyzed in \cite{KR10}. The icosahedral rotation group for example has no invariant subgroup and so cannot 
be a semidirect product of two subgroups.

Eq. \ref{or14} shows that each of the subgroups generates the cosets for its partner.

\subsection{Topology and deck groups for orbifolds.}
\label{Top}

We turn to the topological significance of the fundamental duplex domains for the group $S\Gamma$.
A duplex of a unimodular Coxeter group on the 3-sphere 
cannot  form a topological 3-manifold, because under the point group action it exhibits fixpoints of finite order $k>1$ on its boundaries.  The fixpoints are located on the rotation axis passing through the faces of the duplices, and their order is simply the order of the corresponding rotations. 

The order $k$ of a point is defined as the order of its stabilizer.
Points of order $k=1$ in topology are called regular, for order $k>1$ singular, \cite{RA94} pp. 664-6.
We propose here to move in topology from the standard space forms to  orbifolds. 
This concept is illustrated on the square in Fig. \ref{fig:Fig5}.

We refer to Montesinos \cite{MO87} pp. 78-97 and 
to Ratcliffe \cite{RA94} pp. 652-714 for the introduction and mathematical terms associated with  orbifolds. The notion  includes a manifold structure and a covering, but  
admits singular points of finite order.

We claim: 

{\bf Prop 3: Spherical Platonic 3-manifolds with point symmetry  are appropriately described  as orbifolds.} 

We  shall  demonstrate orbifold coverings by deck transformations from $S\Gamma$, and derive 
the  harmonic analysis for their use in cosmic topology.

In figures \ref{fig:Fig6},  \ref{fig:Fig7}, \ref{fig:Fig8}, \ref{fig:Fig9} we reproduce, with minor changes, based on 
\cite{KR10}, \cite{KR10A}, a set of fundamental duplex domains for the point groups. Now we interprete them as 3-orbifolds under  $S\Gamma$ acting on the 3-sphere, based on  the tetrahedron, the cube, the octahedron or the dodecahedron,  drawn  in their Euclidean version. The edges of the Platonic polyhedra are given in dashed lines, the orbifolds are marked by  yellow color. Note that the polyhedra in the figures are drawn in Euclidean fashion, but stand for spherical polyhedra on $S^3$. 

For topological 3-manifolds, the deck group on the 3-sphere is generated by the operations which map the pre-image
of the manifold to all its face neighbours. We present  a corresponding analysis for the deck group of an
orbifold on the 3-sphere. 

We use the tetrahedral manifold $N2$ from Table \ref{table:Table3}  for the demonstration of orbifold  covering.
Here the Coxeter group is $S(5)$, the symmetric group on $5$ objects, with unimodular subgroup $S\Gamma=A(5)$, the group of even permutations. In Fig. \ref{fig:Fig6} we mark the four vertices of the tetrahedron by the numbers 
$1,2,3,4$. A different  enumeration is used in  \cite{KR10}, Fig. 4.
The Weyl reflection operators are in one-to-one correspondence to the vertex transpositions
\begin{equation}
 \label{or15}
W_1=(1,2),\: W_2=(2,3),\: W_3=(3,4),\: W_4=(4,5).
\end{equation}

From eq. \ref{or15} and Fig. \ref{fig:Fig6}  we find for the four covering deck rotations of this orbifold the simple expressions
\begin{eqnarray}
 \label{or16}
&&(W_1W_3)=(1,2)(3,4),
\\ \nonumber 
&&(W_1W_2)=(1,2)(2,3)=(1,2,3),\: (W_1W_2)^{-1}=(2,3)(1,2)=(3,2,1),
\\ \nonumber 
&&(W_1W_4)=(1,2)(4,5),
\end{eqnarray}
compare Fig. \ref{fig:Fig6}. The three even permutations in eq.  \ref{or16} cover the orbifold and generate $A(5)$.

We return to all four orbifolds associated with Platonic polyhedra and  denote them by $N8, N9$, $N10$ as in \cite{KR10A} and by $N11$ for 
the Platonic dodecahedron.
Each face of the orbifold as proto-tile of the duplex tiling is covered by a face-neighbour, a  copy of the orbifold. 
Instead of drawing the neighbours, we give in the Figures \ref{fig:Fig6}-\ref{fig:Fig9}  the deck rotations from $S\Gamma$ which map the 
preimage into its face neighbours. These deck rotations have axes which intersect with faces of the orbifold. In the figures we denote the rotation axes 
by blue lines and give in each case the rotation as an even product of Weyl reflections. The first three deck rotations 
all have axes passing through the center of the polyhedron and generate  the point group $M$.
The last even product always involves a Weyl reflection that passes through an outer face of the Platonic polyhedron. The  rotation  containing this Weyl reflection
has order $k=2$. It transforms the orbifold into a face neighbour  inside  a  new Platonic neighbour polyhedron.   By examining the deck operations from the figures and comparing them with the Coxeter group relations, it can be verified 
in each case that the four deck operations that cover the faces of the orbifold generate the full unimodular Coxeter group $S\Gamma$. This is in full analogy to the role of  deck groups $H$  for spherical 
3-manifolds \cite{KR10}.

The intersections of the covering  rotation axes with the 3-orbifold determine singular  points and their orders. The center point of the initial spherical polyhedron, chosen as $x=(1,0,0,0)$ and marked in the figures in black, is a fixpoint of maximal order $k=|M|$. For the orbifolds under inspection, more  fixpoints appear on the inner  points of edges of the glue  triangle of each duplex, with area 
marked in the figures in light blue. The order $k$ of these fixpoints agrees with the order of the covering rotation. Note that, in contrast to the Euclidean drawings, we always infer the order $k$ of the rotations 
from the Coxeter group relations and their spherical settings on the 3-sphere.
We give the covering rotations and orders of these  fixpoints in 
Table \ref{table:Table3}.

The obvious rotation axes from the point group $M$ determine additional  singular points 
on inner positions of edges of the orbifolds. For all  four orbifolds we find by inspection:

{\bf Prop 4: Deck transformations of orbifolds generate unimodular Coxeter groups}: Deck rotations  cover  the four faces of the four Platonic 3-orbifolds, they generate the duplex tiling of $S^3$ by all elements of the corresponding unimodular Coxeter deck groups $S\Gamma$. It follows that the orbifolds can abstractly be characterized as quotient spaces $S^3/S\Gamma$.

\section{Harmonic analysis on orbifolds.}
\label{Har}

The harmonic analysis can follow the two views described in section \ref{Eig}.
Clearly the homotopic boundary conditions imply  selection rules compared to  the full basis on the 3-sphere.
The  CMB radiation on an assumed  topology usually is  modelled by random coefficients in the specific  basis.
Here we analyze the basis construction, but postpone any numerical modelling.

On the 3-sphere, after replacing the Cartesian coordinates $x$  by  matrix coordinates $u=u(x) \in SU(2,C)$ \cite{KR10}, the harmonic basis may be spanned by Wigner polynomials $ D^j_{m_1m_2}(u)$ \cite{ED57}. More convenient for pure three-dimensional rotations of 3-space $(x_1, x_2, x_3)$ as they appear in the point groups  are  the  spherical
polynomials $ \psi (jlm)(u)$, linearly related to the Wigner polynomials by 

\begin{eqnarray}
 \label{or17}
&&\psi (jlm)(u)= \sum_{m_1m_2} D^j_{m_1m_2}(u) \langle j-m_1jm_2|lm\rangle (-1)^{(j-m_1)},
\\ \nonumber
&& D^j_{m_1m_2}(u)= \sum_{lm} \psi (jlm)(u)\langle j-m_1jm_2|lm \rangle (-1)^{(j-m_1)},\: 0\leq l\leq 2j,
\end{eqnarray}
and with summations restricted by the Wigner coefficients \cite{ED57}.
The spherical polynomials transform,  under rotations of the last three coordinates $(x_1,x_2,x_3)$ while preserving $x_0$, like the spherical harmonics $Y^l_m$ \cite{KR10},
\begin{equation}
\label{or18}
(T_{(g_l,g_l)}\psi) (jlm)(u)=\psi (jlm)(g_l^{-1}ug_l)= \sum_{m'=-l}^{m'=l} \psi (jlm')(u) D^l_{m',m}(g_l).  
 \end{equation}
They are adapted to the application of rotations from the point group.
Point symmetry selects 
a lowest multipole order $l_0$, compare \cite{KR10} Table 3, \cite{KR10A}.
 The Wigner polynomial basis is suited for the projection
of the subbasis invariant under the deck group $H$.
In the cubic case, the deck group 
of the 3-manifold $N3$ is the quaternion group $H=Q$. Invariant under conjugation with  the cubic point group $O$, it forms a semidirect product $S\Gamma=(Q \times_s O)$.

The harmonic analysis on the four spherical orbifolds $N8, N9, N10, N11$ is given by polynomials invariant under 
their unimodular Coxeter group $S\Gamma$. These and only these polynomials repeat their values on 
any Coxeter duplex from the tiling.

{\bf Prop 5: Harmonic analysis on spherical 3-orbifolds}: The harmonic analysis on a spherical orbifold 
is spanned by harmonic polynomials, invariant under its group of deck transformations $S\Gamma$.

We have seen in Prop. 2 that any element of $S\Gamma$ admits a unique factorization. For the projectors 
to the identity representation of $S\Gamma$ we claim by use of eq. \ref{or14}:

{\bf Prop 6: Factorization of projectors}:  The projector to the identity representation, denoted by $\Gamma_1$, for the unimodular Coxeter group  $S\Gamma$ factorizes into the product of the projectors to the identity representations $\Gamma_1$ for the two subgroups $H, M$, 
\begin{equation}
\label{or2}
P^{\Gamma_1}_{S\Gamma}= P^{\Gamma_1}_{H} P^{\Gamma_1}_{M}=P^{\Gamma_1}_{M} P^{\Gamma_1}_{H}.
\end{equation}

{\em Proof}: In the group operator algebra of the unimodular Coxeter group $S\Gamma$ we have 
from eq. \ref{or14} the factorization
\begin{eqnarray}
 \label{or19}
&& P^{\Gamma_1}_{S\Gamma}=\sum_g T_g= \sum_{h,\: p} T_{h p}= \sum_{p,\: h} T_{p h},\:\: h\in H, p \in M,
\\ \nonumber 
 && =\sum_{h,\: p} T_h T_p= (\sum_h T_h)(\sum_p T_p)=(\sum_p T_p)(\sum_h T_h)=P^{\Gamma_1}_{H}\: P^{\Gamma_1}_{M} =P^{\Gamma_1}_{M} P^{\Gamma_1}_{H}.
\end{eqnarray}
This result greatly simplifies the projection to the identity representation:

The projectors $P^{\Gamma_0}_{H}$  for the group H of deck transformations are given in  \cite{KR10} in the Wigner polynomial basis, see eq. \ref{or17}.

{\bf Prop 7: Orbifolds give sharp multipole selection rules}: $S\Gamma$-invariant polynomials by their point group $M$-invariance select  a lowest non-zero multipole order $l_0$, see \cite{KR10} Table 3. 
This projection  is carried out  in \cite{KR10A} for the spherical cubic 3-orbifold and multipole order 
$0\leq l\leq 8$. The results are reproduced, and in part corrected compared to \cite{KR10A}, in Tables \ref{table:Table4} and \ref{table:Table5}  in explicit multipole order.

We note that the results depend on a convenient choice of the 
coordinates in the spherical harmonics with respect to the symmetry axes of the orbifolds. Under a general rotation of the system of these coordinates one would obtain many additional terms of the  same multipole order $l$.
The function $R_{2j+1,l}(\chi)$ is a Gegenbauer polynomial given in \cite{KR10} eq. (55).
The analysis   can be extended by the method described in section \ref{Rec}.

A similar analysis applies to the Platonic spherical tetrahedral, octahedral and dodecahedral 3-orbifolds. 

\begin{table}[t]
$
 \begin{array}{l|l}
l &Y^{\Gamma_1,l}=\sum_m a_{lm}Y^l_m(\theta, \phi)\\ \hline
0 &                             Y^0_0\\
4 &\frac{1}{2}\sqrt{\frac{7}{3}}Y^4_0+\frac{1}{2}\sqrt{\frac{5}{6}} (Y^4_4+Y^4_{-4})\\
6 &\frac{1}{2\sqrt{2}}          Y^6_0-\frac{\sqrt{7}}{4}            (Y^6_4+Y^6_{-4})\\
8 &\frac{\sqrt{33}}{8}          Y^8_0+\frac{1}{4}\sqrt{\frac{7}{6}} (Y^8_4+Y^8_{-4})
                                     +\frac{1}{8}\sqrt{\frac{65}{6}}(Y^8_8+Y^8_{-8})
\end{array}
$

\caption{\label{table:Table4} The first cubic invariant  spherical harmonics $Y^{\Gamma_1,l}$, expressed by spherical harmonics $Y^l_m(\theta, \phi)$. $\Gamma_1$ denotes the identity representation of the point group $H$.}
\end{table}

\begin{table}[t]
$
 \begin{array}{l|l|l}
2j& l      &\psi^{0,\Gamma_1,2j}=\sum_{l} b_{l} R_{2j+1\; l}(\chi)Y^{\Gamma_1,l}(\theta, \phi)\\ \hline
0 & 0      & R_{1 0} Y^{\Gamma_1, 0}\\
4 & 0,4    & \sqrt{\frac{2}{5}} R_{5 0} Y^{\Gamma_1, 0} + \sqrt{\frac{3}{5}} R_{54} Y^{\Gamma_1, 4}\\
6 & 0,4,6  & \sqrt{\frac{1}{7}} R_{7 0} Y^{\Gamma_1, 0} + \sqrt{\frac{6}{11}}R_{7 4}Y^{\Gamma_1, 4}
                                                      +2\sqrt{\frac{6}{77}}R_{76} Y^{\Gamma_1, 6}\\
8 & 0,4,6,8& \sqrt{\frac{1}{57}}   R_{9 0} Y^{\Gamma_1, 0}
            +9\sqrt{\frac{2}{2717}}R_{9 4} Y^{\Gamma_1, 4}\\
          &&+2\sqrt{\frac{38}{165}}R_{9 6} Y^{\Gamma_1, 6}
            +\sqrt{\frac{2}{1235}} R_{9 8} Y^{\Gamma_1, 8}\\
 \end{array}
$
\caption{\label{table:Table5} Harmonic analysis on the cubic Platonic orbifold $N3$: The first $(S\Gamma=(Q \times_s O))$-invariant polynomials $\psi^{0,\Gamma_1,2j}$
of degree $2j=0,4,6,8$  on the 3-sphere, expressed by  Gegenbauer polynomials $R_{nl}(\chi)$ and cubic invariant  spherical harmonics
$Y^{\Gamma_1,l}$ 
from  Table \ref{table:Table4}. $(Q \times_s O)$-invariance enforces superpositions of several cubic invariant  spherical harmonics.
Note the suppression of odd and $l=2$  multipole orders in the eigenmodes.}
\end{table}

\section{Homotopic boundary conditions from orbifolds.}
\label{Hom}

From \cite{KR10} we know that a fixed geometric shape of a Platonic 3-manifold can have different and inequivalent topologies, characterized by different groups $H$ of homotopies and deck transformations. 
These differences give rise to different homotopic boundary conditions. We now examine the boundary conditions for orbifolds.

We have seen that the orbifold  is covered face-to-face by its rotational  images.
It follows, as in the case of 3-manifolds, that the topology on 3-manifolds implies homotopic boundary conditions on the faces of the 3-orbifolds. Since with the orbifolds  we introduce point symmetry in addition to 
deck transformations, we find from the arguments given in \cite{KR10} for random point symmetry,

{\bf Prop 8: Topological universality from point symmetry}: If we demand, for  a function on a given Platonic 3-manifold, point symmetry under $M$ in addition to the boundary conditions set by homotopy on faces and edges, 
then new boundary conditions apply universally, that is, dependent on the point group $M$ but independent of the specific group $H$ of deck transformations chosen. 

{\bf Prop 9: Universal homotopic boundary conditions from 3-orbifolds}: If in addition to homotopy we demand on the manifold symmetry under the  rotational point group $M$, the homotopic boundary conditions for different deck 
and homotopy groups on the same Platonic geometrical shape coincide with one another and reduce to  the homotopic boundary conditions for the orbifolds. Their boundary conditions are determined by the 
generators in Table \ref{table:Table1} of the covering rotations.

\section{Recursive computation of the bases invariant under the orbifold deck groups $S\Gamma=H\cdot M$.}
\label{Rec}

We describe the recursive construction of $S\Gamma$-invariant bases by use of the factorization
given in Prop. 4. This recursive construction is used to obtain the results of Tables \ref{table:Table4}, \ref{table:Table5}. The multiplicity $m(l,\Gamma_1)$ of the identity representation $\Gamma_1$ of $M$ for given multipole order $l$ is given 
in \cite{LA74} pp. 436-8.

Our basic tool are the relations \cite{KR10} between Wigner  and spherical harmonic polynomials on the 3-sphere given in 
eq. \ref{or17}.

(i) We start from a linear combination of spherical harmonics of multipole order $l$,  invariant under the point group $M$, and construct with  a Gegenbauer polynomial $R_{nl}(\chi)$ and its coefficients $a_{lm}$  a $M$-invariant linear combination of spherical polynomials $\psi_{j l m}(u),\: l\leq 2j$ eq. \ref{or19},  on the 3-sphere \cite{KR10} with lowest degree $l= l_0=2j$, 
\begin{equation}
\label{or20}
\psi_{j,l}^0(u)= \sum_m' a_{lm'}\psi_{j,l,m'}(u).
\end{equation}
By the upper index $0$ we denote invariance under the point group $M$.
The starting linear combinations of spherical harmonics  for lowest multipole order $l_0>0$ for the point groups involved can be found in  the literature. For tetrahedral point symmetry the lowest multipole order from \cite{KR10} Table 3  is $l_0=3$. For cubic point symmetry $M=O$ this function has $l_0=4$ and is given in Table \ref{table:Table4}. 
Since the point group $M$ acts by rotations eq. \ref{or18} only, the coefficients $a_{lm'}$ in eq. \ref{or20} are independent of the degree $n=2j$ of the polynomial.

(ii) Next we transform with eq. \ref{or17} from the spherical  to the Wigner basis and apply the projector $P^{\Gamma_1}_H$   for $H$-invariance,  for example the one with $H=Q$ for the cubic manifold $N3$ given in eq. \ref{or11}. If the resulting  function does not vanish, we 
transform with eq. \ref{or17} 
back to the spherical basis. In this way we find from the starting function a new one,  invariant from Prop. 4 under both $M$ and $H$ and hence under $S\Gamma$, given by 
\begin{eqnarray}
\label{or21}
&&\psi_{j,\Gamma_1}(u)= \sum_{i} \left[\sum_m a^{\Gamma_1}_{l+i, m} \psi_{j,l+i,m}(u)\right], 
\\ \nonumber
&&a^{\Gamma_1}_{l+i, m}
= \sum_{m',m_1',m_2',m_1,m_2}
a_{lm'}  \langle j-m_1'jm_2'|lm'\rangle (-1)^{(j-m_1')}
\\ \nonumber
&&\langle jm_1'm_2'|P^{\Gamma_1}_{H}|j m_1m_2\rangle
\langle j-m_1jm_2|l+i\: m\rangle (-1)^{(j-m_1)}.
\end{eqnarray}
The polynomial eq. \ref{or21}, if non-vanishing, can be normalized.

The Wigner coefficients are given in \cite{ED57}, and the matrix elements of the projector $P^{\Gamma_1}_H$ for the identity representation of the group $H$ are given in the Wigner polynomial basis by
\begin{equation}
\label{or22}
\langle jm_1'm_2'|P^{\Gamma_1}_{H}|j m_1m_2\rangle
=\frac{1}{|H|} \sum_{h =(h_l,h_r)\in H} D^j_{m_1'm_1}(h_l^{-1})D^j_{m_2'm_2}(h_r), 
\end{equation}
and specified in \cite{KR10} for  each Platonic  3-manifold. Due to universality Prop. 5, we can choose the most convenient deck group $H$ for a fixed geometric shape.
The recursion relation eq. \ref{or21} involves  Wigner coefficients, the elements of the group $H$  given as pairs
$h=(h_l, h_r)$ in \cite{KR10}, and Wigner $D^j$-representations for the group $SU(2,C)$. 
The $S\Gamma$-invariant bases appear as linear combinations of $M$-invariant spherical functions
with fixed multipole order $l+i$.

(iii) Moreover, since the point group action cannot change the multipole order,
each new partial sum  of eq. \ref{or21} in square brackets for fixed $l+i$ must generate a (new)  invariant under the point group $M$. This allows to restart the computation with $l \rightarrow l+i$ by going again from spherical to Wigner polynomials, followed by projection of an invariant under $H$. In this way we can increase the polynomial degree $2j$.
In Tables \ref{table:Table4}  and \ref{table:Table5}  we show the first steps in the recursive computation of eigenmodes for the group $Q \times_s  O$.
By character technique, see \cite{KR08}, \cite{KR09}, we can control the number of invariants for given degree $2j$ of the polynomials.

\newpage
\section{Applications.}
\label{Appl}
Before applying the eigenmodes as found in Tables \ref{table:Table4},  \ref{table:Table5} to the analysis of CMB radiation, one has to consider several  questions from astrophysics. In principle 
the CMB radiation spectrum reflects the density fluctuations of the primordial plasma at the time of last scattering. In the cosmological process of creation and propagation  of the 
CMB radiation one has to consider the Sachs-Wolfe and the Doppler effect.  An obstacle  is a strong foreground contamination of the observed radiation due to CMB sources within our own galaxy, see  \cite{WE04}, which the WMAP team tried to eliminate. 
The Ulm group of physicists has experience in incorporating all these  astrophysical data. 
The standard procedure then is to choose a topological manifold, to consider random superpositions of its eigenmodes up to a maximal  polynomial degree and multipole order, to 
incorporate the astrophysical data, and 
compute  spectra similar to  Fig. \ref{fig:Fig1}. 
An application of the cubic Platonic 3-manifold $N3$ that includes all astrophysical data is given  in cooperation with the Ulm group
in \cite{AU11}.

\section{Conclusion.}
In the present analysis, we follow  J Levin in \cite{LE02} p. 305, who in the context of hyperbolic 3-manifolds sees  no convincing reason to omit the study of orbifolds in relation to the CMB. We  illuminate  the topological background of cosmic topology and then  develop new methods of group representations for the analysis of eigenmodes for spherical orbifolds. 
We see the  future interpretation and use of spherical 3-orbifolds from Table \ref{table:Table1} in cosmic topology along two lines:

(I) We can retain  the strict original notion and analysis of topological 3-manifolds. The new  harmonic basis
characterized by point symmetry forms   a   subbasis of the harmonic analysis based on the deck group $H$. If this subbasis   fails to model the CMB fluctuations, 
one can return to  the larger  basis of \cite{KR10} for the 3-manifold without point symmetry.

(II) We formulate the notion of  a topological 3-manifolds with (random) point symmetry $M$ in terms   of spherical 
3-orbifolds. The relevant groups of deck transformations are shown to be four unimodular Coxeter groups $S\Gamma$.
They generate, from a Coxeter duplex orbifold as prototile, a tiling of the 3-sphere into $|S\Gamma|$ copies.
We factorize these groups as $S\Gamma = H \cdot  M $ into products of a group $H$ of polyhedral deck transformations with the point group $M$
of the polyhedron. The Platonic  3-manifolds under the assumption of random point symmetry are replaced by  3-orbifolds, and
their harmonic bases are supported  on a fraction $1/|S\Gamma|$  of the 3-sphere.

For the deck group $S\Gamma= H \cdot M$  we construct a new  harmonic analysis which can model the CMB radiation.
Point symmetry implies topological universality: all fundamental groups
for the same geometrical polyhedral shape produce the same boundary conditions.
In contrast to general spherical space forms, point symmetry enforces, depending on the chosen point symmetry,  a sharp non-zero onset of the multipole expansion of CMB radiation
with harmonic polynomials of minimal non-trivial degree $2j_0=l_0$. 
This phenomenon is in line with the physics of molecules 
with point symmetry \cite{LA74}. The  multipole selection rules, if present, must show up already in the lowest multipole orders of the CMB radiation.
  Moreover, topology correlates polynomials  of increasing multipole order with one another.
These phenomena are  exemplified   in Tables \ref{table:Table4}, \ref{table:Table5}  for orbifolds from spherical cubes.

{\bf Acknowledgement}:  The author is grateful to Tobias Kramer for the  recursive algebraic computation of the results given  in Tables \ref{table:Table4}  and \ref{table:Table5}.

\end{document}